\documentclass[11pt]{article}

\usepackage[preprint]{acl}
\usepackage{times}
\usepackage{latexsym}

\usepackage[T1]{fontenc}

\usepackage{amsmath,amsfonts,bm}

\def\eqref#1{equation~\ref{#1}}

\def\1{\bm{1}}

\DeclareMathAlphabet{\mathsfit}{\encodingdefault}{\sfdefault}{m}{sl}
\SetMathAlphabet{\mathsfit}{bold}{\encodingdefault}{\sfdefault}{bx}{n}

\usepackage{adjustbox}
\usepackage{booktabs}
\usepackage{hyperref}
\usepackage{multirow}
\usepackage{tabularx}
\usepackage{url}
\usepackage{enumitem}

\usepackage{wasysym}
\usepackage{pifont}  
\usepackage{amssymb}   
\usepackage{varwidth}
\usepackage{balance}

\usepackage{tcolorbox}
\usepackage{enumitem} 
\tcbset{
  colback=gray!5,
  colframe=black,
  boxrule=0.5pt,
  arc=2pt,
  left=4pt,
  right=4pt,
  top=4pt,
  bottom=4pt,
  fonttitle=\bfseries,
  before skip=6pt,
  after skip=6pt,
  boxsep=3pt
}

\usepackage{mdframed}
\usepackage{verbatim}
\usepackage{listings}
\usepackage{xcolor}

\renewcommand{\arraystretch}{0.9}

\graphicspath{{./fig/}}

\title{Can LLM Agents Really Debate? \\A Controlled Study of Multi-Agent Debate in Logical Reasoning}

\author{
  Haolun Wu\textsuperscript{1, 2}\thanks{These authors contribute equally to this work.}, 
  Zhenkun Li\textsuperscript{3}\footnotemark[1],
  Lingyao Li\textsuperscript{3}\footnotemark[1]
  \vspace{1mm}
  \\
  \textsuperscript{1}~McGill University,
  \textsuperscript{2}~Mila - Quebec AI Institute,
  \textsuperscript{3}~University of South Florida
  \vspace{1mm}
  \\
  \texttt{haolun.wu@mila.quebec},
  \texttt{\{zhenkun, lingyaol\}@usf.edu}
}

\begin{document}

\maketitle
\begin{abstract}
Multi-agent debate (MAD) has recently emerged as a promising framework for improving the reasoning performance of large language models (LLMs). Yet, whether LLM agents can genuinely engage in deliberative reasoning—beyond simple ensembling or majority voting—remains unclear. We address this question through a controlled study using the \textit{Knight–Knave–Spy} logic puzzle, which enables precise, step-wise evaluation of debate outcomes and processes under verifiable ground truth. We systematically setup six structural and cognitive factors, including agent team size, composition, confidence visibility, debate order, debate depth, and task difficulty, to disentangle their respective effects on collective reasoning. Our results show that intrinsic reasoning strength and group diversity are the dominant drivers of debate success, while structural parameters such as order or confidence visibility offer limited gains. Beyond outcomes, process-level analyses identify key behavioral patterns: majority pressure suppresses independent correction, effective teams overturn incorrect consensus, and rational, validity-aligned reasoning most strongly predicts improvement. These findings provide valuable insights into \textit{how} and \textit{why} LLM debates succeed or fail, offering guidance for designing interpretable and truth-seeking multi-agent reasoning systems.
Our dataset and code are available at \href{https://anonymous.4open.science/r/ControlMAD-CFE7/README.md}{this link}.
\end{abstract}

\section{Introduction}
\label{sec:intro}


In real-world problem-solving, debate enables groups to combine knowledge, cross-check reasoning, and correct errors~\citep{branham2013debate}. Recent studies show that large language models (LLMs) can achieve similar benefits through debate~\citep{estornell2024multi, liang2024encouraging, chan2023chateval, liu2025truth}. When multiple agents critique and refine each other's answers, they often reach higher accuracy than a single LLM, supporting the ``society of minds'' view~\citep{du2023improving}, with further demonstrated gains in mathematics~\citep{zhang2025debate4math}, healthcare decision-making~\citep{lu2024triageagent}, and factual reasoning~\citep{du2023improving}. 
Debate can also mitigate hallucinations and logical fallacies while producing interpretable dialogue traces~\citep{duan2025enhancing, lin2024interpreting, ma2025debate}.



However, challenges remain about the nature of LLM debate. One key argument is whether the reported gains reflect genuine debate
or simply the effects of ensembling and majority voting~\citep{zhang2025if}. In addition, bias amplification and echo chambers are concerns~\citep{oh2025understanding, estornell2024multi}: when agents share similar training or biases, debates can reinforce incorrect beliefs rather than challenge them~\citep{liu2025breaking}. In particular, agents do not always revise their stance when confronted with correct counterarguments, while an eloquent but incorrect agent can sometimes sway others ~\citep{agarwal2025persuasion}. These limitations raise a key question: \textbf{Can LLM agents really debate to advance their understanding?} 
To address this question meaningfully, we argue that a controlled study must examine both the debate outcome (i.e., whether accuracy improves) and the process (i.e., whether agents engage in rational interaction).
Therefore, we ask two questions.

\begin{itemize}[leftmargin=1.3em, itemsep=0pt, topsep=1pt, parsep=1pt, partopsep=1pt]
    \item \textbf{RQ1 (Outcome). What factors influence debate outcomes?} How do factors such as self-reported confidence, player order, agent heterogeneity, debate depth, and the presence of strong and weak reasoners significantly influence debate dynamics and solution outcomes?
    \item \textbf{RQ2 (Process). How do agents engage in effective debate processes?} To what extent do LLM agents actually engage in meaningful debate, such as identifying mistakes, adopting peer suggestions, or revising their answers?
\end{itemize}



To answer these questions, we adopt the \textit{Knight--Knave--Spy} logic puzzles as a controlled reasoning environment that enables systematic measurement of both debate outcomes and debate processes. 
We design a multi-agent debate framework that simulates realistic deliberation: each agent makes initial proposals, debates others’ reasoning, and revises its beliefs step-by-step. 
By systematically varying numerous factors, we quantify how each influences collective reasoning accuracy. 
Beyond outcomes, we further propose three desiderata for effective debate and empirically analyze how these properties correlate with performance gains. 
Together, these experiments provide a controlled lens on how debate structure and reasoning behavior jointly determine success or failure in multi-agent debate systems on logical reasoning.
Our main contributions are as follows:

\begin{itemize}
    \item We present a benchmark based on the \textit{Knight--Knave--Spy} puzzle that enables rigorous evaluation of multi-agent debate on logical reasoning under verifiable ground truth.

    \item Through controlled experiments, we systematically analyze how debate design choices affect debate outcomes, revealing that intrinsic model strength is the dominant factor governing debate success.

    \item We move beyond outcome accuracy by introducing desiderata of effective debate and analyzing the debate process, showing that process-level behaviors aligned with these properties correspond to higher correction rates and collective reasoning performance.

\end{itemize}


\section{Related Work}
\label{sec:related} 

Multi-agent debate (MAD) addresses the limitations of single LLMs in multi-step reasoning, including hallucinations and ``degeneration-of-thought'' ~\citep{chan2023chateval, du2023improving, li2024improving, wang2025learning, su2025debflow}. By structuring adversarial and cooperative communication among agents, MAD has been shown to improve performance across tasks, such as mathematical reasoning~\citep{zhang2025debate4math}, fact checking~\citep{kim2024can, he2025debating}, healthcare decision-making~\citep{lu2024triageagent, kim2025tiered}, and code summarization~\citep{chun2025multi}. Debate frameworks like Society of Minds~\citep{du2023improving} and ChatEval~\citep{chan2023chateval} demonstrate how debate enhances factual grounding, interpretability, and robustness. 

To make debates effective, much existing research has focused on improving the outcome of debate by developing structured protocols that coordinate how agents contribute and deliberate. Judge-based systems employ a designated expert to select the most convincing argument~\citep{liang2024encouraging, khan2024debating}, while consensus-driven approaches use majority voting or iterative convergence~\citep{kaesberg2025voting, li2024improving}. Debate can also be enriched by agent heterogeneity, where prompt variations, diverse models, or role assignments are utilized to encourage diverse argumentation and reduce blind spots~\citep{smit2024mad, ye2025x, yang2025revisiting, xing2025designing}. Another strategy is confidence reporting, where agents provide answers with self-assessed confidence scores~\citep{lin2025enhancing, bai2024confidencecal, eo2025debate}. Last, prior work highlights that debate depth and turn-taking strategies can influence outcomes~\citep{lu2024triageagent, chan2023chateval}.

Despite its promise, LLM debate faces fundamental challenges that call its efficacy into question~\citep{choi2025debate, oh2025understanding, zhang2025if}. A primary concern is whether observed performance gains arise from interactive argumentation or are simply the aggregation of multiple independent outputs~\citep{oh2025understanding, estornell2024multi}. Persuasiveness can eclipse accuracy, enabling eloquent but incorrect arguments to prevail over sound reasoning. LLM judges can compound this problem through systematic biases such as verbosity, positional preference, and sycophancy~\citep{khan2024debating}. Homogeneous groups risk amplifying shared biases, creating echo chambers~\citep{bandaru2025revealing} or premature consensus around incorrect solutions~\citep{kaesberg2025voting}. Moreover, extended debate depth does not always improve outcomes~\citep{ku2025multi}; additional rounds can entrench initial errors or foster groupthink. These dynamics raise doubts about whether debate can effectively foster truth-seeking.

This body of work reveals a critical gap: while the potential benefits and failure modes of agent debate are known, many of the underlying interaction dynamics remain unmeasured. It is therefore difficult to distinguish meaningful, process-driven reasoning from simple outcome aggregation. Our study addresses this gap through a set of controlled experiments to unpack not only the factors that drive debate outcomes but also evaluate the effectiveness of the debate process itself, shedding light on \textit{how} and \textit{why} MAD succeeds or fails.


\section{Task and Dataset: \textit{Knight-Knave-Spy}}
\label{sec:task}

We situate our work in the \emph{Knight--Knave--Spy} puzzle, a classic logic game for evaluating deductive reasoning.
Each \emph{player} is assigned one of three roles: a \emph{knight} (always tells the truth), a \emph{knave} (always lies), or a \emph{spy} (may either tells the truth or lie).
Players make natural-language statements about themselves or others, and the objective for the solving \emph{agents} is to infer the correct role assignment for every player consistent with all game constraints.

We choose this game as the testbed for MAD for three reasons:
(\romannumeral1) \textbf{Reasoning challenge.} Solving these puzzles requires nontrivial logical reasoning and consistency checking, making them challenging even for human solvers.
This ensures that LLM agents must engage in genuine reasoning rather than rely on shallow heuristics.
(\romannumeral2) \textbf{Stepwise structure.} The debate progresses player by player: agents focus on one player's statement at a time, evaluate consistency with others, and decide whether to revise their judgment.
This granularity is important for isolating how intermediate decisions are influenced during debate, unlike other reasoning tasks (e.g., long math word problems or open-domain question answering) where reasoning steps are implicit or under-specified.
(\romannumeral3) \textbf{Clear evaluation.} The task yields unambiguous ground-truth solutions, enabling precise measurement of accuracy, error cascades, and adoption dynamics.
In contrast, tasks such as summarization or commonsense reasoning often admit multiple valid solutions, making evaluation noisier and interpretation less clear.
Together, these properties make the \emph{Knight--Knave--Spy} puzzle a controlled yet demanding environment for studying \textit{when} and \textit{why} MAD succeeds or fails.

We construct a dataset spanning puzzle sizes from 4 to 9 players, with 300 distinct cases per size, totaling 1,800 puzzles.
This setting enables probing scaling effects: smaller games involve simpler reasoning, while larger ones increase difficulty and cascade risk.
An example with four players is illustrated below.
\begin{tcolorbox}[title=Game size-4\, id-1, colback=white, colframe=black!50, fonttitle=\small\bfseries]
\small 
\textbf{Player statements}  
\begin{itemize}[left=0pt,itemsep=1pt,topsep=1pt]
    \item Rachel: ``Violet and I have the same role.''  
    \item Violet: ``Rachel is telling the truth.''  
    \item Olivia: ``Among Violet and Rachel, exactly one person is telling the truth.''  
    \item Peter: ``Among the following two statements, exactly one is true: (1) Among all players, the number of knaves is even. (2) Among Rachel, Violet, and Olivia, the number of people who are lying is odd.''  
\end{itemize}

\vspace{0.5mm}
\textbf{Game manager hint:}\\ Among all players, there is exactly one spy.  

\vspace{1.0mm}
    \textbf{Solution:} 
Rachel = knight; Violet = knight; Olivia = knave; Peter = spy  
\end{tcolorbox}

\section{Multi-Agent Debate Framework}
\label{sec:method}

\begin{figure*}[t]
    \centering
    \includegraphics[width=1.0\linewidth]{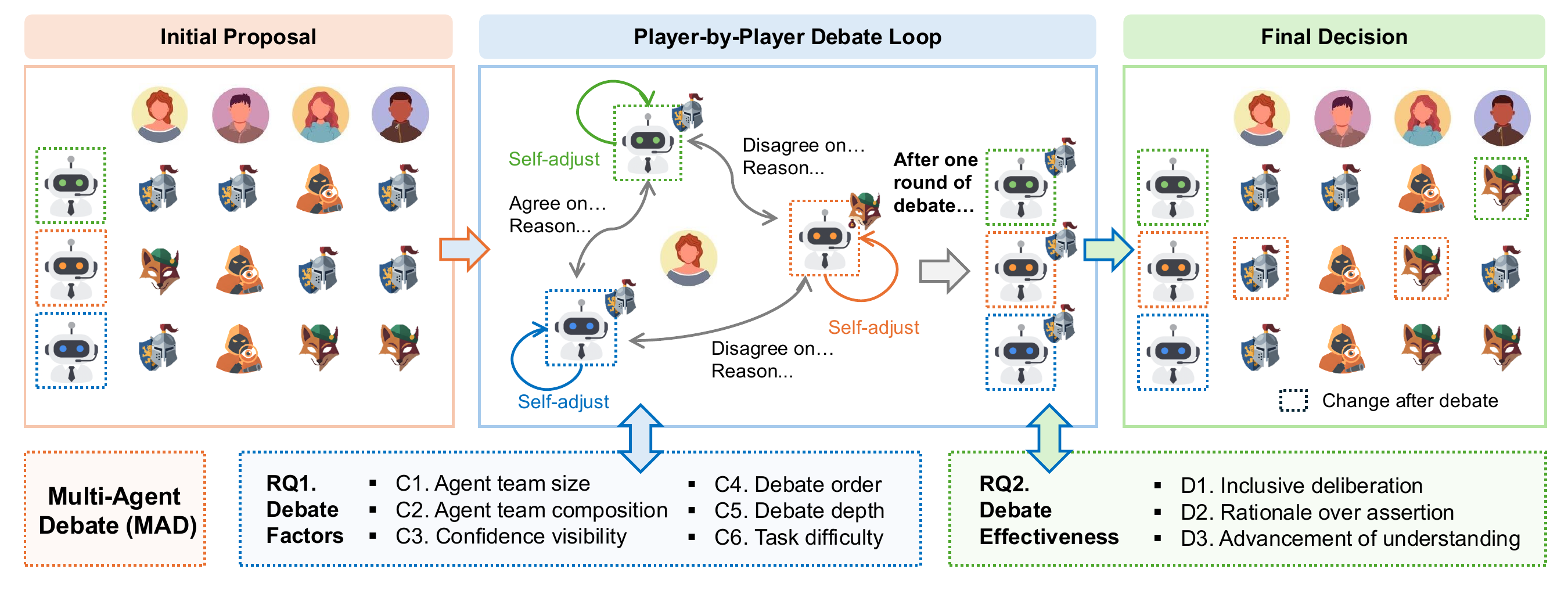}
    \vspace{-10mm}
    \caption{The illustration of the overall MAD framework design.}
    \vspace{-2mm}
    \label{fig:framework}
\end{figure*}

Motivated by prior agent debate studies~\cite{liang2024encouraging, chan2023chateval, li2024improving}, we design a framework that structures MAD into explicit phases aligned with the stepwise nature of the \textit{Knight--Knave--Spy} puzzle.  
Each player's statement defines an intermediate step, allowing LLM agents to reason, debate, and revise the role at a time before reaching a final joint solution.  
The system maintains separate chat histories for each agent, ensuring self-awareness.  

\subsection{Debate Protocol}\label{sec: debate protocol}
As prior studies often includes initial proposal, debate, and self-reflection processes~\cite{liang2024encouraging, chan2023chateval, li2024improving}, the entire debate protocol is designed as follows (see Figure~\ref{fig:framework}). The prompt design for each phase is detailed in Appendix~\ref{app:prompt}.

\begin{itemize}[leftmargin=1.5em, itemsep=1pt, topsep=1pt, parsep=2pt, partopsep=1pt]
    \item \textbf{Initial proposal.}  
    Each agent independently assigns roles to \emph{all players} and reports a confidence score for each assignment.  
    This establishes a baseline set of beliefs for later debate.

    \item \textbf{Player-by-player debate loop.}  
    The core of the framework is a structured loop over all players.  
    For each player $i$, two sub-phases are executed in sequence: (i) \emph{Debate phase:} Agents take turns presenting arguments about player $i$'s role, referencing both the player's statement and logical consistency with prior claims. They may agree or disagree with peers and justify their reasoning. (ii) \emph{Self-adjustment phase:} After hearing all arguments about player $i$, each agent reviews the discussion and decides whether to revise its own label for that player. This loop ensures that every player is discussed in depth and that revisions are localized to one role at a time.
    \item \textbf{Final decision.}  
    Once all players have been debated and revised, each agent outputs its complete set of role assignments.  
    The system then aggregates these outputs through majority voting on a \emph{per-player} basis.  
    If no consensus is reached for a player, a designated supervisor agent (e.g., \textit{gpt-5}) breaks ties.
\end{itemize}

\subsection{Factors Shaping Debate Outcomes}
\label{sec:control_factor}

The design of a multi-agent debate can vary in many structural and cognitive aspects, each influencing how effectively agents reason and reach consensus. 
To systematically understand the effect of different factors on debate outcomes, we thoroughly review prior studies~\citep{bai2024confidencecal, lu2024triageagent, chan2023chateval, smit2024mad, hegazy2024diversity, liu2024groupdebate} and identify six controllable factors, as summarized in Table~\ref{tab:controllable-factors}. 
Each factor leads to a dedicated controlled experiment (C1–C6) that quantifies its impact on debate outcomes, with detailed setups presented in Section~\ref{sec:control_exp}.
This examination provides a foundation for understanding how design choices shape collective reasoning success (aligned with RQ1).

\begin{table*}[t]
\centering
\small
\renewcommand{\arraystretch}{1.1} 
\caption{Controllable factors in the MAD framework (Section~\ref{sec:control_factor}) and their corresponding controlled experiments (Section~\ref{sec:control_exp}).  
Each factor is varied independently relative to the default anchor configuration (\textbf{A}), detailed in Section~\ref{sec:control_exp}, to isolate its effect on debate outcomes.}
\vspace{-3mm}
\label{tab:controllable-factors}
\resizebox{\linewidth}{!}{
\begin{tabular}{p{3.0cm}p{5.8cm}p{6.2cm}}
\toprule
\textbf{Factor} & \textbf{Explanation} & \textbf{Controlled Experiment (relative to anchor A)} \\
\midrule

\textbf{Agent team size} &
Number of debating agents may affect the diversity of perspectives and the stability of consensus. &
\textbf{C1:} Increase team size from 3 to 4 by adding the strongest agent.  
Tests whether larger groups improve debate accuracy. 
\\

\textbf{Agent team composition} &
Team diversity may affect reasoning complementarity.  
Homogeneous teams repeat one model's reasoning; heterogeneous teams mix different strengths and styles. &
\textbf{C2:} Compare homogeneous vs.\ heterogeneous teams.  
Evaluates whether model diversity improves reasoning. \\

\textbf{Confidence visibility} &
Visible self-reported confidence may speed convergence but risk overconfidence cascades. &
\textbf{C3:} Allow agents to see each other's confidence scores.  
Assesses whether visibility stabilizes or destabilizes debate. \\

\textbf{Debate order} &
The order of player discussion may shape how early judgments influence later reasoning. &
\textbf{C4:} Let agents agree on the debate order instead of using a fixed one.  
Tests if different ordering influences outcomes. \\

\textbf{Debate depth} &
Multiple passes allow more reflection but may yield diminishing returns. &
\textbf{C5:} Increase depth from one to two full passes.  
Measures whether extended deliberation improves accuracy. \\

\textbf{Task difficulty} &
Puzzle size determines reasoning complexity: larger games require longer inference chains. &
\textbf{C6:} Vary game size from 5 to 9 players.  
Examines how task complexity scales debate performance. \\
\bottomrule
\end{tabular}
}
\vspace{-3mm}
\end{table*}

\subsection{Desiderata of Effective Debate Processes}
\label{sec:desiderata}

Beyond measuring final accuracy, it is equally important to examine the process through which debates unfold. 
Outcome-based evaluations reveal \emph{how} performance changes, but they do not explain \emph{why} certain debates lead to genuine understanding while others fail. 
To uncover these underlying mechanisms, we analyze the debate process itself and propose three desiderata that characterize the essential qualities of effective reasoning and interaction (aligned with RQ2).

\noindent
\textbf{$\blacklozenge$
 Inclusive deliberation.}
An effective debate requires that participants consider and respond to each other's arguments rather than speaking past one another, showing engagement by acknowledging, incorporating, or critically examining peers' reasoning instead of ignoring input or repeating claims. Moreover, minority views should not be automatically silenced by majority pressure; inclusivity ensures that all participants have the opportunity to contribute and their reasoning is given consideration even when it does not align with the dominant position. Observable signals include agents revising positions when presented with valid counterarguments and minority-held correct answers being adopted rather than suppressed.

\noindent
\textbf{$\blacklozenge$
 Rationale over assertion.}
An effective debate prioritizes reasoning grounded in evidence rather than unsupported claims.
Participants should justify their positions with verifiable information and logical consistency, not merely assert conclusions.
This manifests when position changes correlate with argument validity rather than rhetorical confidence or consensus frequency. 

\noindent
\textbf{$\blacklozenge$
 Advancement of understanding.}
Finally, an effective debate should leave participants and observers with improved clarity or insight.
A debate that merely reinforces pre-existing biases or entrenched positions, without generating new understanding, cannot be considered successful.
Concretely, agents should correct initial errors through reasoning exchange, not merely aggregate independent guesses.

We note that effective debate desiderata are context-dependent. In this work, we focus on \emph{logical reasoning tasks with verifiable ground truth}, where evidence-grounding, attentive engagement, and accuracy improvement form the core of constructive deliberation.  
We present detailed empirical analysis in Section~\ref{sec:results_rq2}.

\section{Experiment Design}
\label{sec:experiment}

In this section, we first describe how agent teams are constructed and define the default anchor configuration. We then outline the controlled experiments, each varying a single factor to measure their effect in a systematic manner. Measurements for both outcome accuracy and reasoning dynamics provide a comprehensive evaluation.

\begin{figure}[t]
    \centering
    \includegraphics[width=1.0\linewidth]{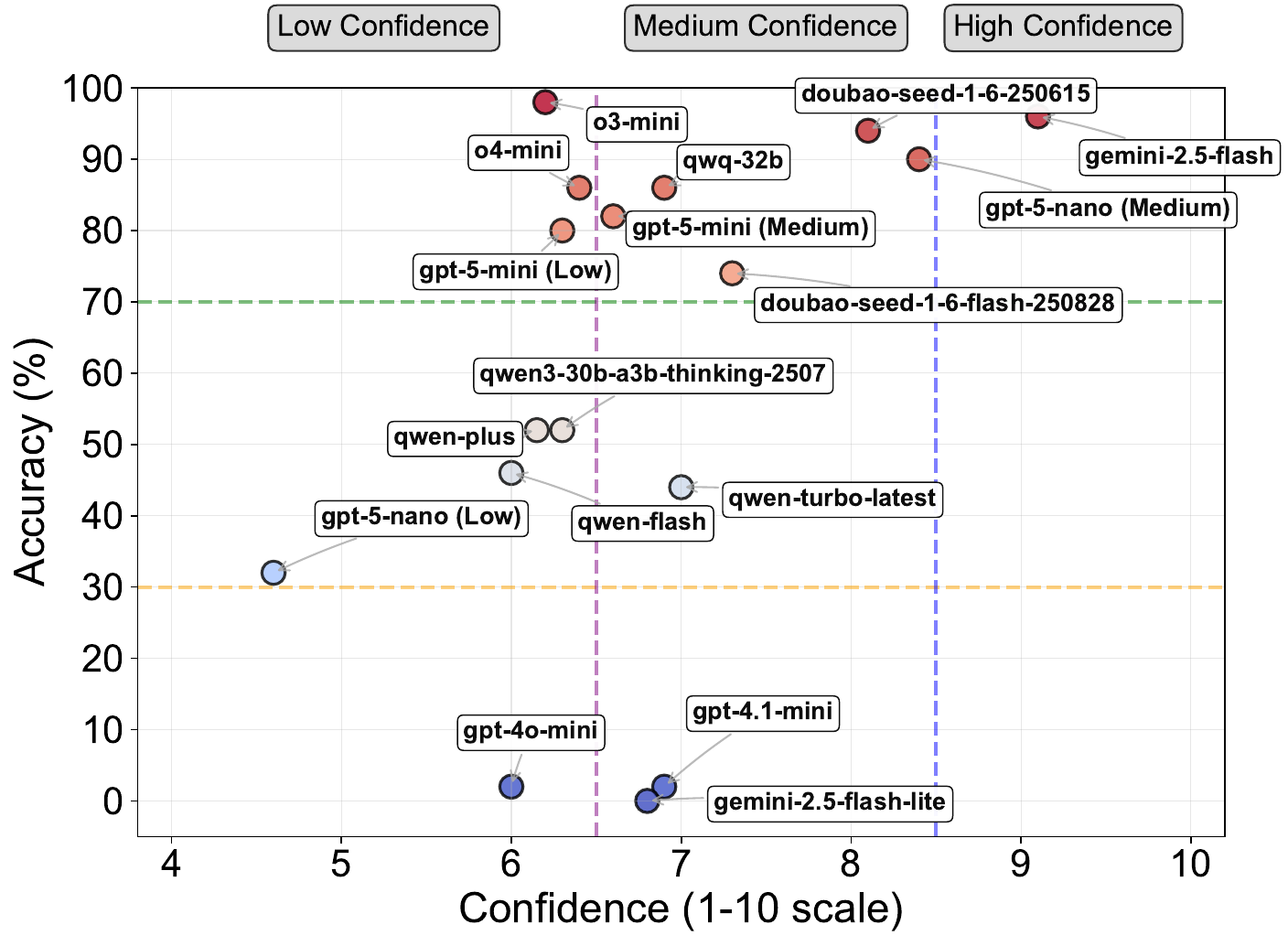}
    \vspace{-6mm}
    \caption{Single-agent accuracy versus self-reported confidence over 100 medium-difficulty games (size=6). 
This benchmark guides the agent team formation by categorizing models into high / medium / low performance and confidence.
}
\vspace{-3mm}
    \label{fig:accuracy-confidence}
\end{figure}

\begin{table*}[t]
\centering
\small
\caption{Agent team configurations used in our experiments.
Each team consists of three agents unless otherwise specified.
Performance (Perf.) and confidence (Conf.) levels correspond to single-agent accuracy and self-reported confidence shown in Figure~\ref{fig:accuracy-confidence}.
Each dimension has three levels: $\blacktriangle$~High, $\CIRCLE$~Medium, and $\blacktriangledown$~Low.
}
\vspace{-3mm}
\label{tab:team-compositions}
\begin{tabular}{lll}
\toprule
\textbf{Team Setting} & \textbf{(Perf., Conf.) Combination} & \textbf{Agent Team Formation} \\
\midrule
\textbf{Het-Mix A (balance)} & 
($\blacktriangle$, $\CIRCLE$), ($\CIRCLE$, $\CIRCLE$), ($\blacktriangledown$, $\CIRCLE$) &
\textit{gpt-5-nano (Medium)}, \textit{qwen-turbo-latest}, \textit{gemini-2.5-flash-lite} \\

\textbf{Het-Mix B (stress)} &
($\blacktriangle$, $\blacktriangledown$), ($\CIRCLE$, $\blacktriangledown$), ($\blacktriangledown$, $\CIRCLE$) &
\textit{gpt-5-mini (Low)}, \textit{qwen-flash}, \textit{gemini-2.5-flash-lite} \\

\textbf{Het-Mix C (diversity)} &
($\blacktriangle$, $\blacktriangle$), ($\CIRCLE$, $\blacktriangledown$), ($\blacktriangledown$, $\CIRCLE$) &
\textit{gemini-2.5-flash}, \textit{qwen-flash}, \textit{gpt-4.1-mini} \\

\textbf{Het-Mix D (strong)} &
($\blacktriangle$, $\blacktriangle$), ($\blacktriangle$, $\CIRCLE$), ($\blacktriangle$, $\blacktriangledown$) &
\textit{gemini-2.5-flash}, \textit{gpt-5-nano (Medium)}, \textit{gpt-5-mini (Low)} \\

\textbf{Hom-Mix Strong} &
($\blacktriangle$, -) × 3 &
\textit{gpt-5-nano (Medium)} × 3 \\

\textbf{Hom-Mix Weak} &
($\blacktriangledown$, -) × 3 &
\textit{gpt-4.1-mini} × 3 \\
\bottomrule
\end{tabular}
\vspace{-3mm}
\end{table*}

\subsection{Agent Team Construction}

We construct multiple agent teams to explore how different reasoning and confidence profiles affect debate performance.  
Each individual model is first mapped into the accuracy–confidence space (Figure~\ref{fig:accuracy-confidence}),  
where accuracy is averaged over 100 medium-difficulty games (size=6), and confidence refers to the model's self-reported score on a 1–10 scale.  
Models are categorized along two axes:  
(i) \textbf{Performance (Perf.)}, measured by accuracy, and  
(ii) \textbf{Confidence (Conf.)}, based on self-reported scores,  
each with three levels: $\blacktriangle$~High, $\CIRCLE$~Medium, and $\blacktriangledown$~Low.
This categorization yields a $3\times3$ grid of possible agent types.

Table~\ref{tab:team-compositions} summarizes the six various team settings.  
These configurations cover a broad range of reasoning strengths and confidence profiles.  
\textbf{Het-Mix A} balances strong and weak agents with moderate confidence.  
\textbf{Het-Mix B} stress-tests groups containing overconfident weak agents,  
\textbf{Het-Mix C} maximizes diversity across both dimensions, and  
\textbf{Het-Mix D} explores the upper bound of performance with mostly strong agents.  
The two homogeneous settings (\textbf{Hom-Mix Strong} and \textbf{Hom-Mix Weak}) provide baselines without model diversity,  
allowing comparison of heterogeneous versus homogeneous reasoning dynamics.

\subsection{Default Anchor and Controlled Settings}
\label{sec:control_exp}
We establish a default anchor configuration and systematically vary individual factors (C1–C6) to isolate their effects. In particular, we establish a \textbf{default anchor configuration (A)} that serves as the baseline for all controlled experiments.  
This configuration adopts the \textbf{Het-Mix A} team composition as its agent setup,  
while keeping all structural parameters fixed as follows: 
\vspace{-1mm}
\[
\boxed{%
  \begin{varwidth}{1.0\linewidth}
    \textbf{A: } Agent team size = \textit{3}, \;
    Agent team composition = \textit{Het-Mix A}, \;
    Confidence visibility = \textit{hidden}, \;
    Debate order = \textit{fixed}, \;
    Debate depth = \textit{1}, \;
    Task difficulty = \textit{all levels}.
  \end{varwidth}%
}
\]
\vspace{-1mm}
Building on this anchor, we define six controlled experiments (\textbf{C1–C6}), 
each modifying a single factor while keeping all others constant.  
This design enables precise attribution of performance changes to individual debate parameters.  
Table~\ref{tab:controllable-factors} lists the six factors and their corresponding experimental settings.

All experiments are conducted on 100 puzzles per configuration, covering easy, medium, and hard conditions.  
Thus, the analysis of \textit{Task difficulty} is naturally embedded within the evaluation.

\subsection{Metrics and Measures}
\label{sec:metrics}

We evaluate the MAD framework using two complementary groups of measurements: 
(1) \emph{outcome-level metrics} for \textbf{RQ1}, which quantify how different debate configurations affect debating performance, and 
(2) \emph{process-level measurements} for \textbf{RQ2}, which assess whether the debates themselves exhibit desirable reasoning dynamics.

\paragraph{Outcome-level metrics.}
We assess debate performance using outcome-based metrics that capture \emph{how} correctness and consensus evolve throughout the debate. 
These metrics provide a quantitative view of collective reasoning effectiveness by tracking both the accuracy of solutions and the stability of convergence over rounds. 
Specifically, we measure strict and smooth accuracies to evaluate overall correctness at different granularities: strict accuracy counts an instance as correct only when all roles are inferred accurately, whereas smooth accuracy measures the proportion of correctly identified roles within each instance. 
To assess temporal dynamics, we compute their area-under-curve (AUC) variants, which summarize how rapidly and consistently agents approach the correct solution across debate turns. 
Beyond accuracy, we evaluate alignment among agents using AUC agreement under both unanimous and majority criteria, capturing whether discussions lead to coherent and sustained consensus. 
Together, these outcome-based metrics reveal \emph{how} reasoning performance develops over time and how stable collective judgments emerge during the debate. 
Detailed definitions and formulas are provided in Appendix~\ref{appendix:metrics}.

\paragraph{Process-level measurements.}
While outcome-based metrics capture how debates unfold, process-level measurements aim to uncover why they succeed or fail. Motivated by the desiderata introduced in Section 4.3, these analyses move beyond correctness to examine the internal reasoning and interaction patterns that underlie collective improvement. We track agents’ behavior across debate rounds by observing transitions such as when incorrect majorities are overturned or minority-held correct answers are adopted, indicating inclusive deliberation and exchange of reasoning. We further estimate rationality-based correction rates using an external judge model to assess whether agents revise positions in response to valid arguments rather than social conformity. Together, these measurements operationalize the desiderata and explain why some debates lead to deeper understanding while others stagnate.

\begin{figure}[t]
    \centering
    \includegraphics[width=0.46\linewidth]{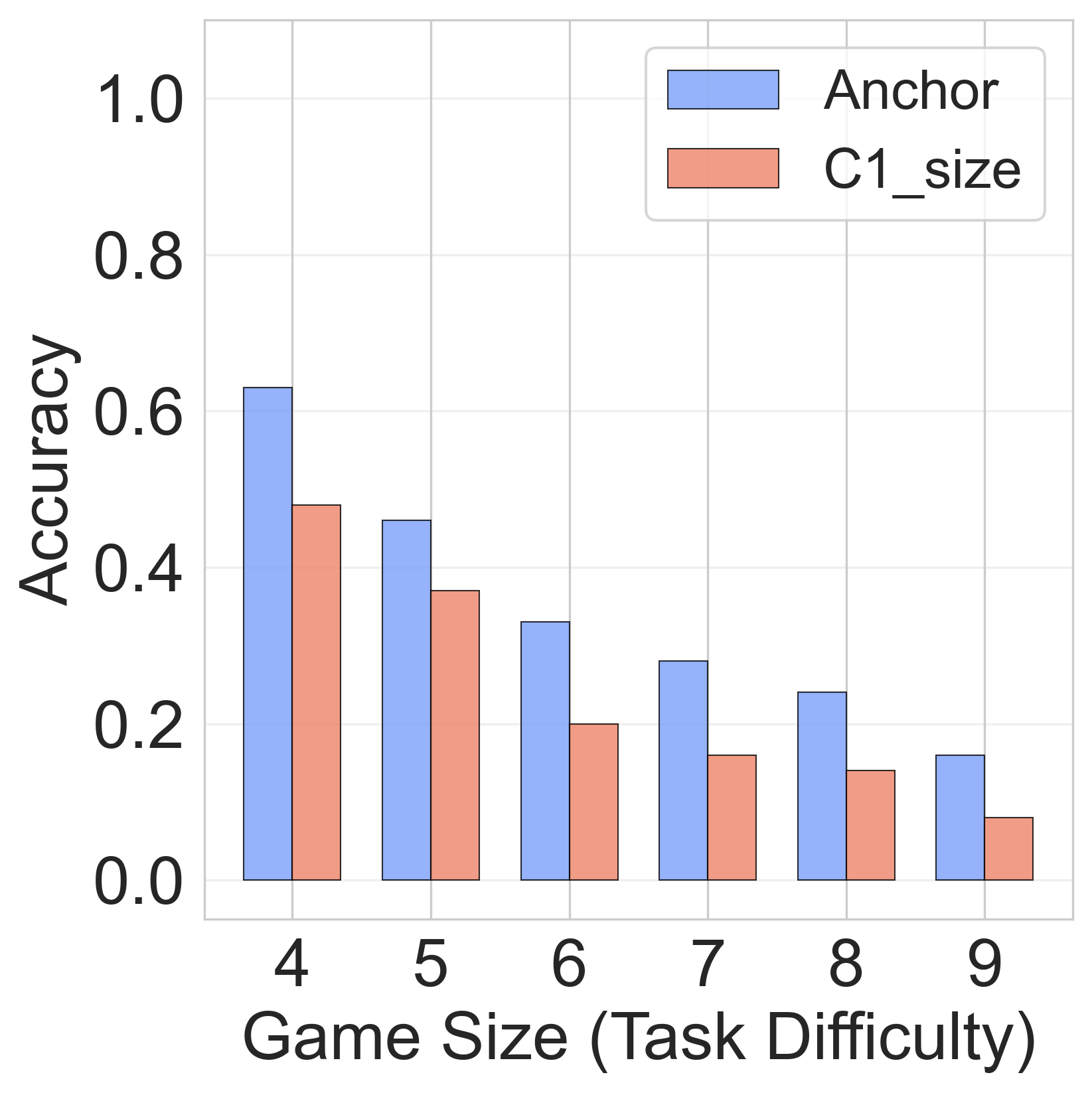}
    \hfill
    \includegraphics[width=0.46\linewidth]{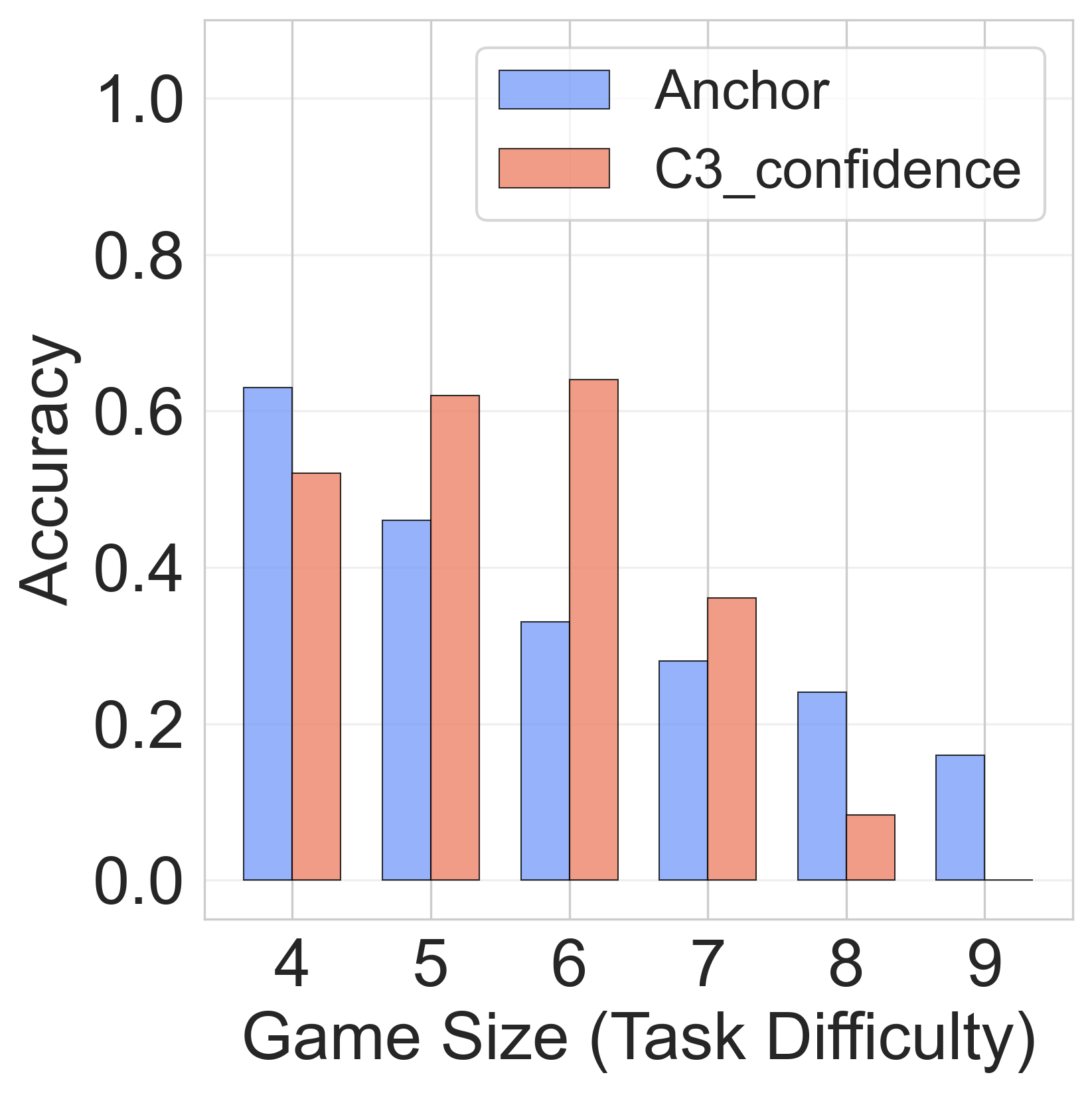}
    \vspace{1em}

    \includegraphics[width=0.46\linewidth]{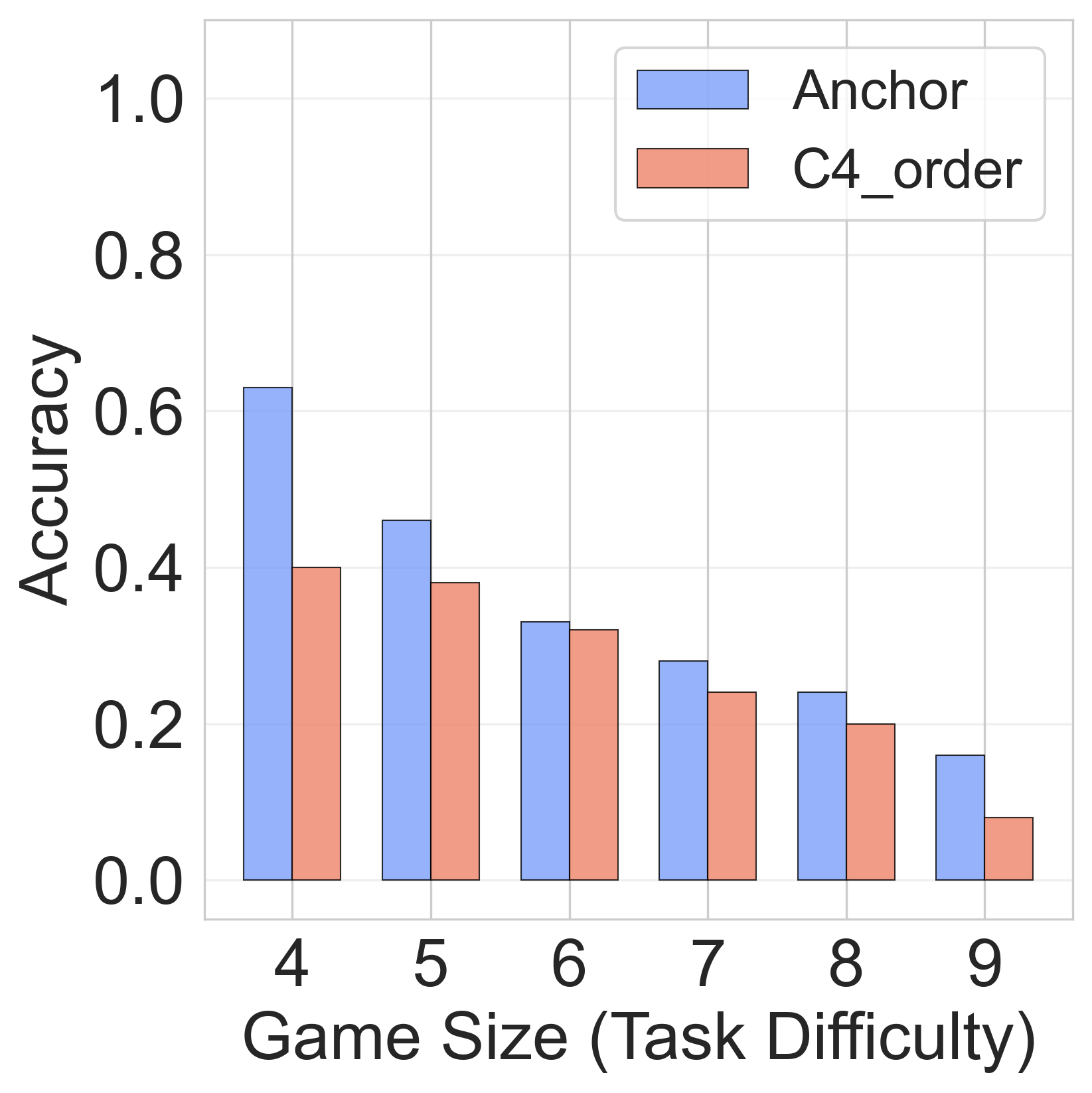}
    \hfill
    \includegraphics[width=0.46\linewidth]{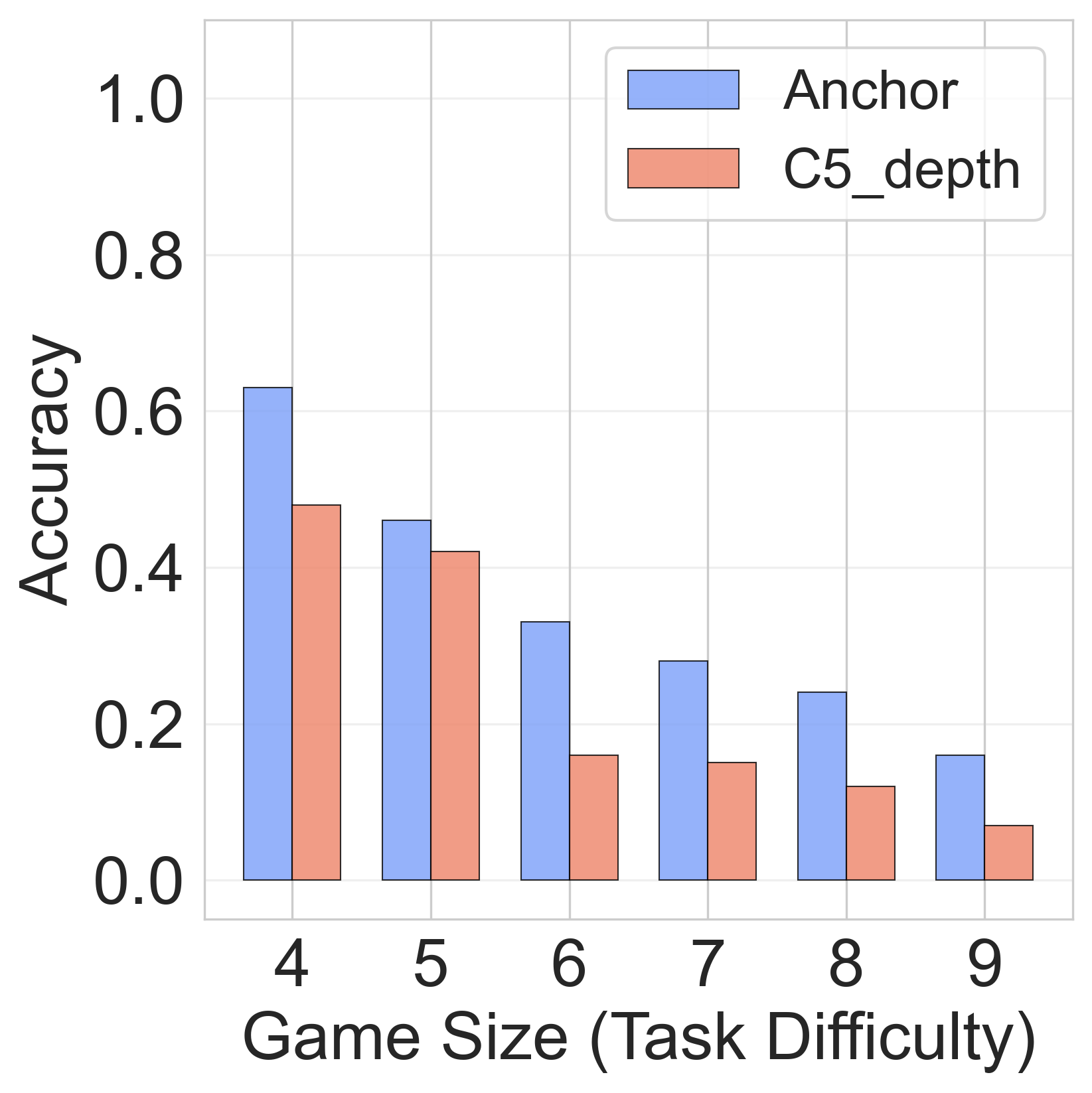}
    \vspace{1em}

    \includegraphics[width=1.0\linewidth]{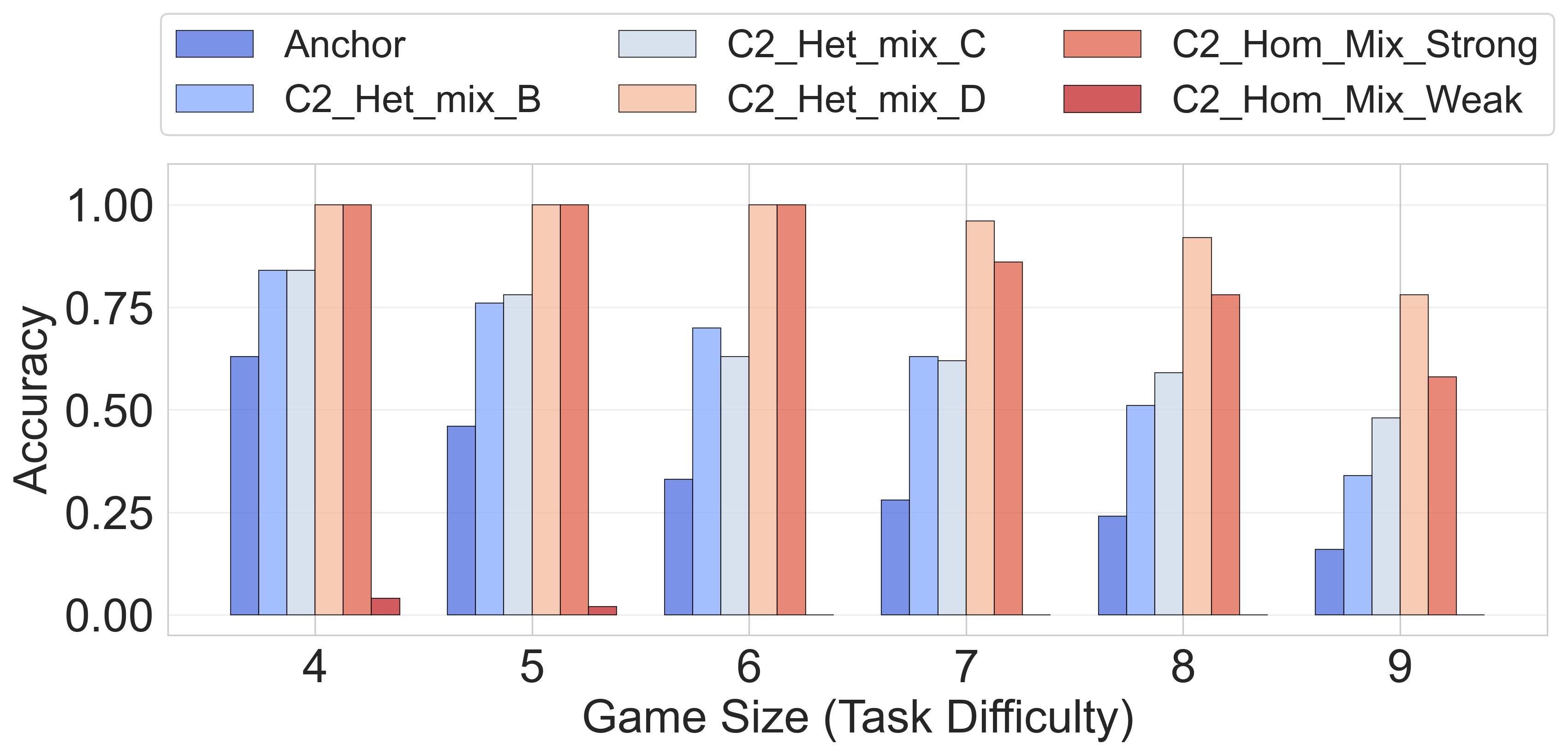}
\vspace{-8mm}
    \caption{Overall accuracy across controlled debate settings relative to the default anchor (A). 
Each panel isolates one factor while keeping others fixed. 
The Task difficulty factor, reflected by varying game sizes, is naturally incorporated in all subfigures.}
    \vspace{-3mm}
    \label{fig:control_results}
\end{figure}

\begin{table}[t]
\centering
\small
\caption{Comparison of initial and post-debate accuracy across game sizes. Debate consistently improves both instance- and agent-level performance, with larger gains in simpler tasks.}\label{tab: initial vs final accuracy}
\vspace{-3mm}
\begin{tabular}{lccc}
\toprule
\textbf{Metric} & \textbf{Size 4} & \textbf{Size 6} & \textbf{Size 8} \\
\midrule
Initial Instance (\%) & 17.33 & 6.00 & 3.33 \\
Final Instance (\%)   & 69.33 & 46.67 & 36.00 \\
\textbf{Improvement (\%)} & \textbf{+52.00} & \textbf{+40.67} & \textbf{+32.67} \\
\midrule
Initial Agent (\%)    & 35.33 & 27.11 & 22.66 \\
Final Agent (\%)      & 77.11 & 55.56 & 44.89 \\
\textbf{Improvement (\%)} & \textbf{+41.78} & \textbf{+28.45} & \textbf{+22.23} \\
\bottomrule
\end{tabular}
\vspace{-4mm}
\end{table}

\begin{table}[t]
\centering
\small
\caption{Regression analysis of factors predicting final output accuracy. Higher initial accuracy, larger teams, and moderate initial chaos enhance performance, while greater task difficulty reduces it.}\label{tab:reg_final_smooth}
\vspace{-3mm}
\begin{tabular}{lccc}
\toprule
\textbf{Variable} & \textbf{Coef.} & \textbf{Std. Err.} & \textbf{Sig.} \\
\midrule
Constant & 0.044 & 0.013 & ** \\
Game size & $-0.030$ & 0.007 & *** \\
\# of agents & 0.066 & 0.014 & *** \\
Debate depth & 0.019 & 0.015 &  \\
Init. smooth acc. & 0.600 & 0.100 & *** \\
Init. strict acc. & $-0.015$ & 0.048 &  \\
Strong init. smooth acc. & 0.170 & 0.091 &  \\
Strong init. strict acc. & 0.005 & 0.057 &  \\
Weak init. smooth acc. & 0.018 & 0.087 &  \\
Weak init. strict acc. & 0.020 & 0.069 &  \\
Strong in minority & 0.011 & 0.012 &  \\
Weak in minority & 0.003 & 0.014 &  \\
Conf. visible (T) & 0.023 & 0.017 &  \\
Debate order same (T) & 0.035 & 0.019 &  \\
Init. has chaos (T) & 0.085 & 0.020 & *** \\
\midrule
$R^2$ & \multicolumn{3}{c}{0.393} \\
Adj. $R^2$ & \multicolumn{3}{c}{0.380} \\
Observations & \multicolumn{3}{c}{745} \\
\bottomrule
\end{tabular}
\begin{flushleft}
\footnotesize
\textit{Note.} Significance levels: * $p<0.05$, ** $p<0.01$, *** $p<0.001$.  
Coefficients represent OLS estimates. Variables marked ``(T)'' are binary indicators (True = 1).  
\end{flushleft}
\vspace{-3mm}
\end{table}

\begin{table}[t]
\centering
\small
\caption{Regression analysis of factors predicting agreement stability. Balanced team composition and moderate variability foster consensus, whereas excessive initial chaos or imbalance hinder it.}\label{tab:reg_auc_all}
\vspace{-3mm}
\begin{tabular}{lccc}
\toprule
\textbf{Variable} & \textbf{Coef.} & \textbf{Std. Err.} & \textbf{Sig.} \\
\midrule
Constant & 0.040 & 0.012 & ** \\
Game size & 0.052 & 0.006 & *** \\
\# of agents & 0.111 & 0.012 & *** \\
Debate depth & 0.019 & 0.013 &  \\
Init. smooth acc. & 0.275 & 0.087 & ** \\
Init. strict acc. & 0.018 & 0.042 &  \\
Strong init. smooth acc. & $-0.022$ & 0.079 &  \\
Strong init. strict acc. & $-0.025$ & 0.049 &  \\
Weak init. smooth acc. & $-0.032$ & 0.076 &  \\
Weak init. strict acc. & 0.067 & 0.060 &  \\
Strong in minority & $-0.102$ & 0.010 & *** \\
Weak in minority & $-0.055$ & 0.013 & *** \\
Confidence visible (T) & 0.025 & 0.015 &  \\
Debate order same (T) & 0.002 & 0.017 &  \\
Initial has chaos (T) & $-0.117$ & 0.017 & *** \\
\midrule
$R^2$ & \multicolumn{3}{c}{0.290} \\
Adj. $R^2$ & \multicolumn{3}{c}{0.274} \\
Observations & \multicolumn{3}{c}{745} \\
\bottomrule
\end{tabular}
\begin{flushleft}
\footnotesize
\textit{Note.} Significance levels: * $p<0.05$, ** $p<0.01$, *** $p<0.001$.  
Coefficients represent OLS estimates. Variables marked ``(T)'' are binary indicators (True = 1).  
\end{flushleft}
\vspace{-3mm}
\end{table}

\begin{figure}[t]
    \centering
    \includegraphics[width=1.0\linewidth]{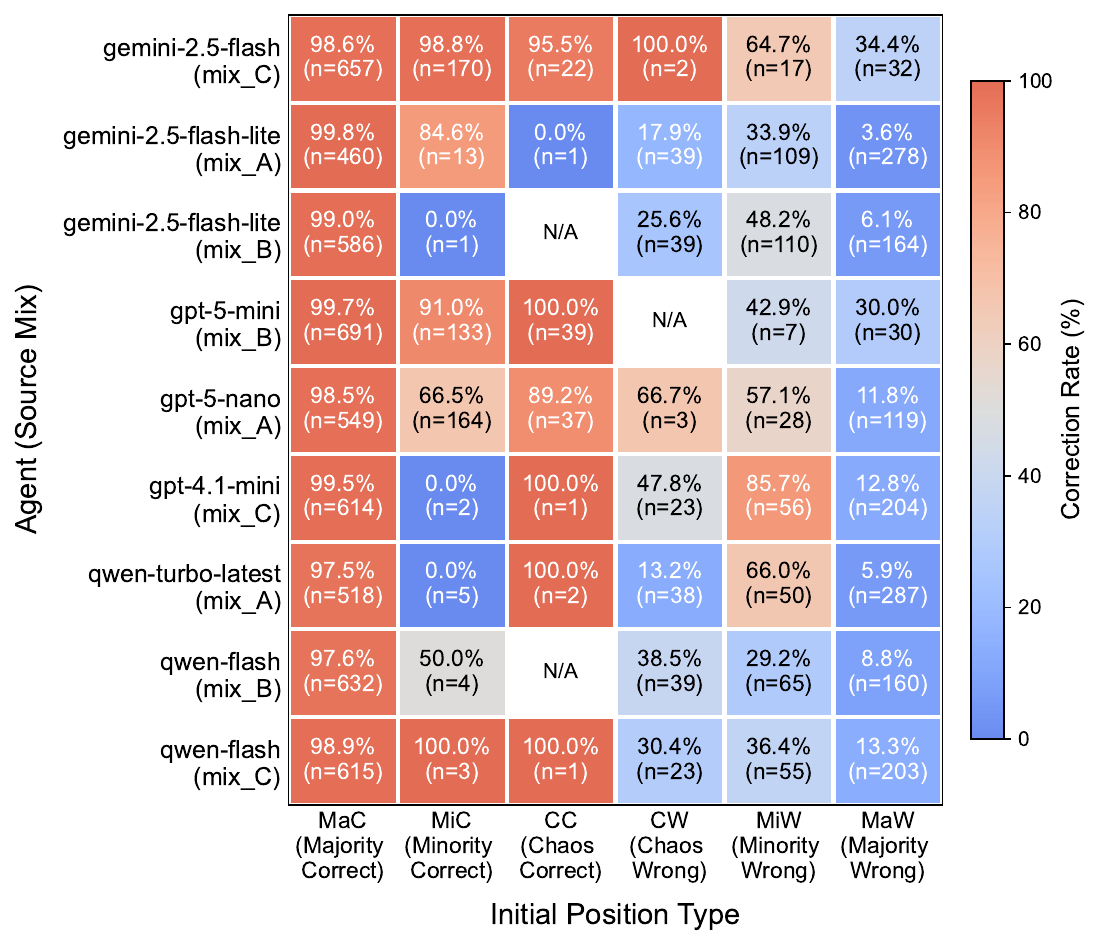}
    \vspace{-8mm}
    \caption{Heatmap on initial states. x-axis: initial stats. Here Ma = majority, Mi = Minorty, first C = chaos, second C = correct, and W = wrong. So MaC means the model starts with a majority and correct initial position in a debate round. y-axis: different models. This figure counts for each agent their initial position before each debate round, and if their final position after the debate round is correct or not (the correction rate).}
    \vspace{-3mm}
    \label{fig:mad-heatmap}
\end{figure}

\begin{figure}[t]
    \centering
    \includegraphics[width=1.0\linewidth]{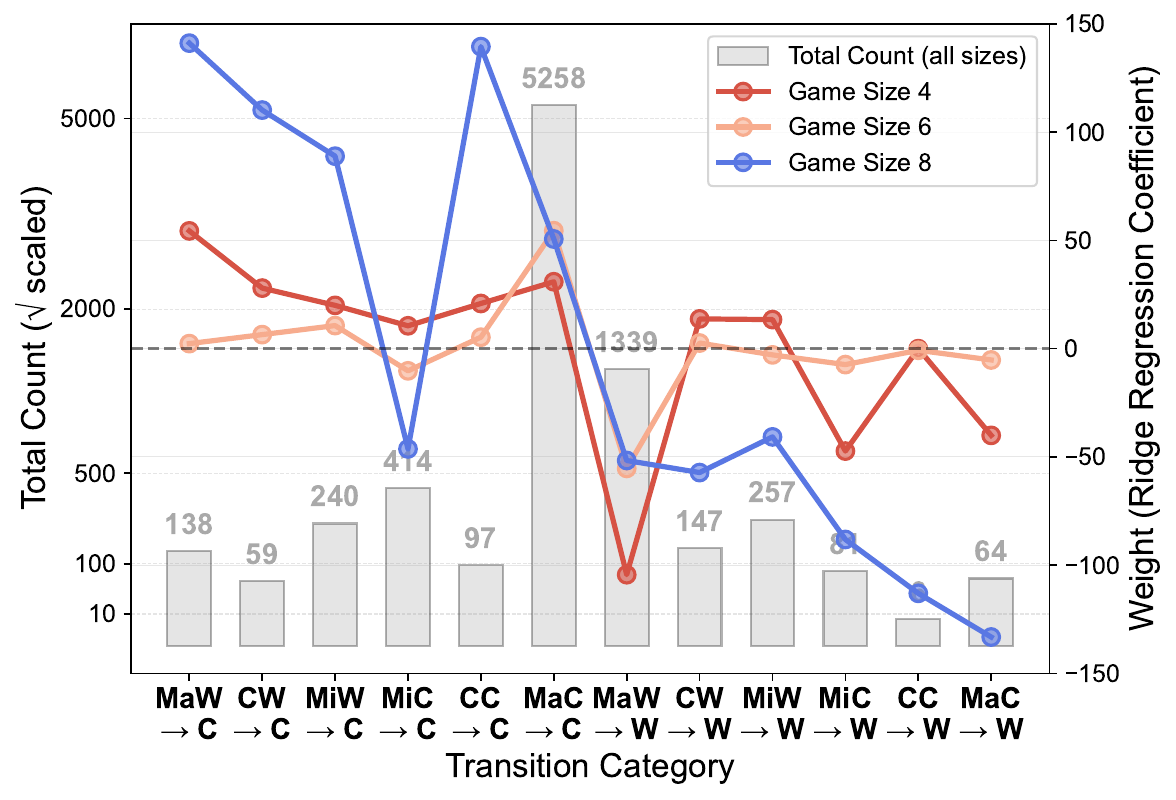}
    \vspace{-8mm}
    \caption{Correlations between states transition and final accuracy of the game instance. x-axis: 12 state transition possibilities. Left y-axis: counts of state transitions. Right y-axis, the weight each state transition contributes to the final accuracy of a game instance: for each game instance we calculate the percentage of players having correct role deduction, not a single 0/1 accuracy for the whole answer. We then employ linear regression methods to simulates the weights and present them in the figure. Higher weights means a state transition contributes positively to the final (smooth) accuracy.}
    \vspace{-5mm}
    \label{fig:mad-performance}
\end{figure}

\begin{figure*}[t]
    \centering
    \includegraphics[width=1.0\linewidth]{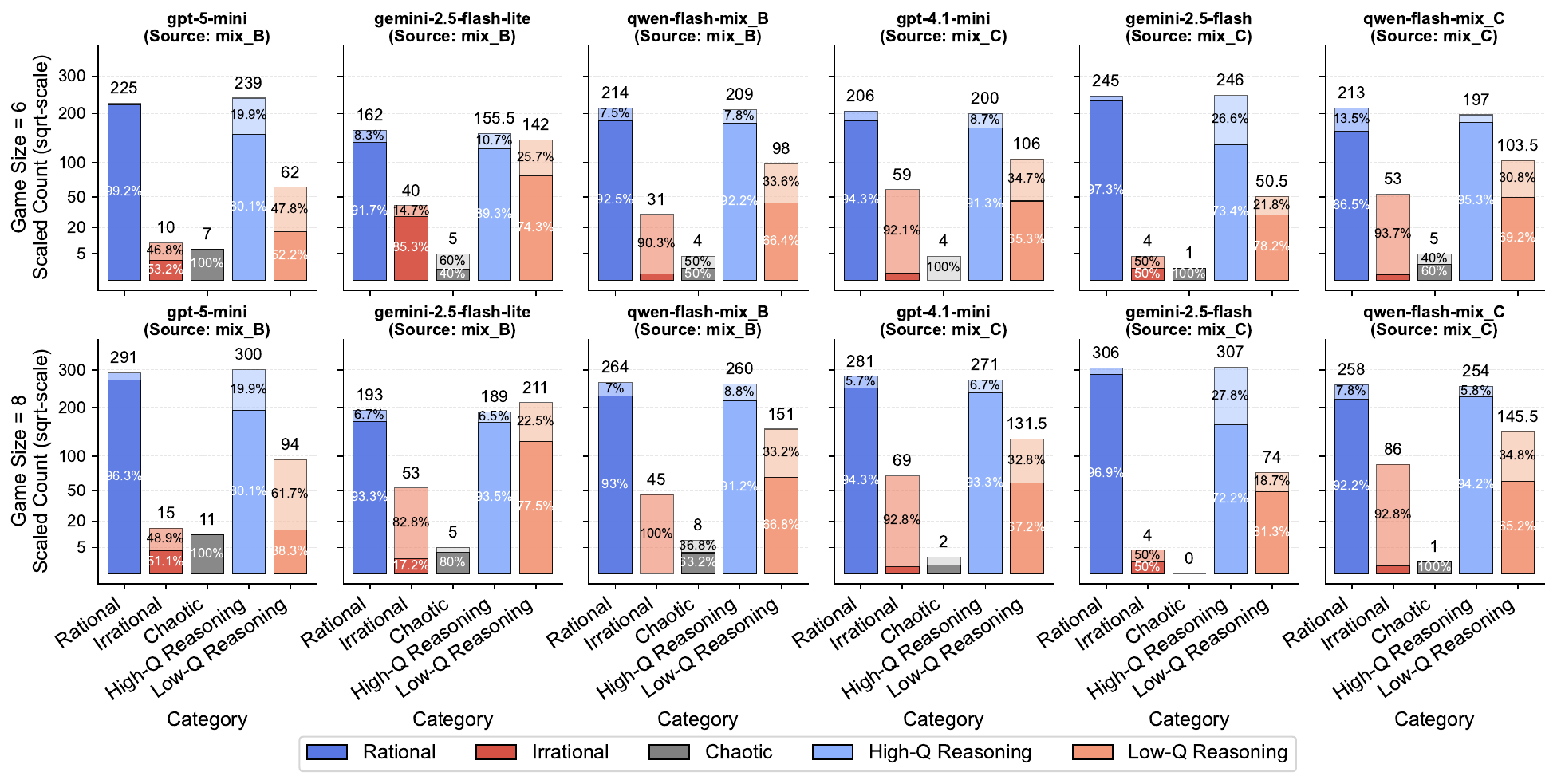}
    \vspace{-6mm}
    \caption{Whether agents behaves rationally. The reasonings in the debate round for each agent is sent to deepseek-r1-0528 model to evaluate (1 through 4). Then we count (in the first and second bar) if agent behaves rationally or not: whether agent follows high (3 or 4) rating opinions from either himself or other agents, as well as whether the final result is correct or not. The third bar counts if the agent has some self-contradicting behaviors. For bar 4 and 5, we cound whether an agents provides a high quality opinions, and correction rate is calculated based on receivers not providers. In the figure, black percentages are failure rates and white ones are the correction rates.}
    \vspace{-4mm}
    \label{fig:mad-reasoning}
\end{figure*}

\section{Results and Analysis}
\label{sec:results}

We now discuss the experiment results corresponding to our two research questions. 

\subsection{RQ1. What Factors Influence Debate Outcomes?} 
\label{sec:results_rq1}

\paragraph{Debate becomes harder as task difficulty increases.}
Figure~\ref{fig:control_results} summarizes the outcomes of all controlled settings relative to the default anchor configuration.  
Across all controlled settings in Figure~\ref{fig:control_results}, performance declines consistently as puzzle size grows.
Larger games require longer inference chains and amplify error propagation, leading to reduced collective accuracy and agreement.
This pattern holds across all debate configurations, indicating that increasing task difficulty uniformly challenges both strong and weak teams.

\paragraph{Debate success is primarily driven by reasoning strength and diversity.}
Direct comparisons of agent team compositions (Figure~\ref{fig:control_results}, bottom panel) show that base model strength governs debate performance. 
\textit{Hom-Mix D (strong)} consistently surpasses \textit{Hom-Mix Weak} at all difficulty levels, establishing a strength-dependent ceiling. 
Among heterogeneous teams, \textit{Het-Mix D (strong)} attains the highest accuracies, followed by \textit{Het-Mix A (balance)}; mixes that concentrate weaker or overconfident agents (e.g., \textit{Het-Mix B (stress)}) lag behind. 
Comparing \textit{Hom-Mix Strong} with \textit{Het-Mix D} indicates that diversity provides modest but consistent gains in stability and accuracy when strong reasoners are present.
By contrast, when all agents are weak (\textit{Hom-Mix Weak}), changing order, depth, or confidence visibility yields negligible benefit. 
Overall, performance remains bounded by the strongest reasoner available, and while diversity provides modest gains, it cannot make up for a team composed entirely of weak agents.

\paragraph{Debate consistently raises collective accuracy within each task.}
As part of our outcome-level evaluation (RQ1), Table~\ref{tab: initial vs final accuracy} compares performance before and after debate across different game sizes.
Instance-level accuracy—measured by majority voting per player and verified against ground truth—shows substantial improvement, increasing by 52\% for size-4 puzzles and 32\% for size-8 puzzles.
A similar upward trend is observed at the agent level, indicating that individual participants also benefit from collective deliberation.
These findings confirm that the debate mechanism itself leads to measurable performance gains within each task, even as overall accuracy remains bounded by task complexity and individual reasoning strength.



To further examine which factors most strongly predict outcome metrics,
we perform regression analyses on \textit{Final Output Accuracy (Smooth)} and \textit{AUC-Agreement-Major}.
The observations on other metrics are similar.
In addition to the six controlled factors defined in Section~\ref{sec:control_factor},
we include a derived variable, \textit{initial chaos}, which captures variability in agents’
first-round predictions before debate begins.
A game instance is labeled as chaotic when no clear majority exists in the initial role assignments,
reflecting early disagreement that may influence subsequent deliberation.
Table~\ref{tab:reg_final_smooth} (accuracy) and Table~\ref{tab:reg_auc_all} (agreement)
summarize the results.
Overall, \textit{initial accuracy}, \textit{agent count}, and moderate \textit{initial chaos}
positively contribute to performance,
whereas increasing \textit{task difficulty} and unbalanced \textit{agent team composition}
reduce agreement stability.

\paragraph{Initial accuracy and team size are the strongest predictors of debate success.}
The regression on \textit{final smooth accuracy} (Table~\ref{tab:reg_final_smooth}) explains
about 39\% of the variance (\(R^2 = 0.393\)), indicating a solid model fit.
The most influential predictor is \textit{initial smooth accuracy} (\(p < 0.001\)),
with a large positive coefficient (\(\beta = 0.600\)):
teams starting from stronger baselines achieve higher post-debate accuracy.
The \textit{number of agents} also has a significant positive effect (\(p < 0.001\)),
suggesting that larger collectives promote improved aggregate reasoning.
In contrast, \textit{game size} shows a negative association (\(p < 0.001\)),
confirming that increasing task complexity hinders accuracy.
Other individual-level variables and majority/minority indicators are not significant,
implying that performance gains arise mainly from baseline quality and group scale
rather than agent positioning.

\paragraph{Moderate initial variability stimulates productive adaptation.}
The presence of \textit{initial chaos} (\(p < 0.001\)) increases final accuracy,
indicating that some early disagreement can promote effective deliberation.
Debate depth, confidence visibility, and debate order remain insignificant,
underscoring the robustness of the core structural factors.
Although detailed mixture coefficients are omitted,
several heterogeneous settings (e.g., Mix~B and Mix~C) show positive effects,
highlighting that team composition also contributes to accuracy improvements.

\paragraph{Team composition and initial balance shape consensus quality.}
The regression on \textit{AUC agreement} (Table~\ref{tab:reg_auc_all}) explains
approximately 29\% of the variance (\(R^2 = 0.290\)), reflecting a moderate fit.
\textit{Game size} (\(p < 0.001\)) and \textit{number of agents} (\(p < 0.001\))
positively predict agreement, showing that larger, more populated debates
reach steadier consensus.
Similarly, \textit{initial smooth accuracy} (\(p < 0.01\)) predicts higher agreement levels.
However, when strong or weak agents begin in minority positions,
agreement quality declines significantly (\(p < 0.001\)),
indicating that balanced influence within teams supports stable convergence.
Conversely, excessive \textit{initial chaos} (\(p < 0.001\)) lowers agreement,
suggesting that while moderate variability aids adaptation,
too much early disorder hinders group alignment.
As before, balanced heterogeneous mixes (e.g., Mix~C, Mix~H)
yield higher agreement scores, reinforcing that diversity, when structured, is beneficial.

\subsection{RQ2. How Do Agents Engage in Effective Debate Processes?}
\label{sec:results_rq2}

\paragraph{Majority pressure suppresses agents’ independent correction.}
We first analyze agent behavior based on their initial position (Figure~\ref{fig:mad-heatmap}) to test the ``Inclusive deliberation'' dimension for debate process. When an agent starts within a correct majority (MaC), its position is exceptionally stable, with all models achieving over 97\% correction rates. However, facing an incorrect majority (MaW) exposes a clear performance gap. Stronger models like \textit{gemini-2.5-flash} and \textit{gpt-5-mini} show a moderate ability to correct the incorrect consensus (34.4\% and 30.0\% correction, respectively), while weaker models are almost entirely swayed by the group, with \textit{gemini-2.5-flash-lite }(mix\_A) correcting itself in only 3.6\% of such cases. This suggests weaker agents can defer to consensus rather than engaging with evidence.

Further investigation reveals a ``minority correction asymmetry'': agents more readily maintain a correct minority position (MiC) than they correct an incorrect one (MiW), indicating a bias toward their initial stance when not under majority pressure. Additionally, we observe clear ``teammate effects,'' where agents like \textit{gemini-2.5-flash-lite} show significantly different correction rates depending on the experimental mix they are in. This confirms that the debate process is not only a function of individual capability but is path dependent and contingent on agent team composition.



\paragraph{Debate success depends on agents’ ability to overturn incorrect consensus.}
We analyze how agents transition between belief states and how these transitions influence overall accuracy (Figure~\ref{fig:mad-performance}). Using ridge regression, we estimate how each transition type affects instance-level smooth accuracy. The distribution of transition counts (gray bars) shows that debates involve stable or consensus-driven states (e.g., MaC$\rightarrow$C, CC$\rightarrow$C), whereas majority-reversal transitions (MaW$\rightarrow$C) are relatively rare. Despite their lower frequency, these reversals exhibit the strongest positive coefficients, indicating that correcting a wrong consensus contributes most to the final accuracy.

Across increasing difficulty levels (red, orange, and blue curves for game sizes 4, 6, and 8), the positive weight of MaW$\rightarrow$C transitions rises sharply, while others like MaC$\rightarrow$W show increasingly negative effects. This pattern suggests that effective debates rely less on maintaining majority agreement and more on agents’ willingness to revise incorrect consensus through minority-informed reasoning. In other words, genuine progress emerges not from conformity but from reflective self-correction when faced with meaningful counterarguments.


\paragraph{Rational, validity-aligned reasoning drives accurate correction.}
We assess debate process by evaluating agents by whether they follows the rationales themselves or provide high quality opinions for others given four setups (game size 6 and 8 in mix B and C) to \textit{deepseek-r1-0528} for soundness rating (1–4, where 3–4 indicates high quality) (Figure~\ref{fig:mad-reasoning}). For agent's own behavior, agents' rationality---correctly following own high rating agree opinion or processing peer arguments and aligning position changes accordingly---strongly predicts success. Rational behaviors yield high correction rates (all above 90\%), while irrational behaviors result in dramatically lower rates (less than 55\%, despite \textit{gemini-2.5-flash-lite}), demonstrating that receptiveness to valid counterarguments is critical for error correction.

As opinion providers, a strong model paradox emerges: when strong models \textit{gpt-5-mini (Medium)} provide high-quality reasoning, their teammates achieve lower correction rates (80\% in game size 8) compared to weaker model providers like \textit{gpt-4.1-mini} (93\%). This reveals that strong agents have better ability in providing and following high-quality opinions. Combined with agent team composition results (Figure~\ref{fig:control_results}), this suggests a key failure mode: weaker models lack capacity to assess opinion quality, leading to less rational behavior that undermines the debate process.



\section{Discussion and Conclusion}
\label{sec:conclusion}
Our findings provide a controlled view of how LLMs engage in MAD for logical reasoning. Debate performance is largely governed by the intrinsic reasoning capability of the participating models: stronger agents consistently lead to more accurate and stable outcomes, while structural factors such as team size, debate depth, or confidence visibility exert only limited influence. These results indicate that coordination mechanisms alone cannot overcome weak reasoning foundations, and that the ceiling of debate success is effectively bounded by the strongest participant.

Beyond aggregate accuracy, our process-level analysis reveals that successful debates are marked by inclusive and rationale-driven exchanges in which agents critically engage with one another, correct errors, and refine shared understanding. In contrast, weaker teams often converge prematurely or follow persuasive yet unsound reasoning. These observations highlight the importance of reasoning diversity, structured argumentation, and feedback mechanisms that promote truth-seeking collaboration. They offer actionable guidance for developing more interpretable, reliable, and self-correcting multi-agent reasoning systems.


\section{Limitations}
Our study has several limitations that suggest directions for future work.
First, we focus exclusively on the Knight--Knave--Spy logic puzzle, a structured domain with unambiguous ground truth. While this choice enables precise measurement of debate dynamics, it may not fully capture the complexities of open-ended reasoning tasks such as creative problem-solving, commonsense reasoning, or tasks with multiple valid solutions. The stepwise nature of our task also imposes a particular debate structure that may not generalize to less structured domains.

Second, our analysis is limited to the LLMs available at the time of study. As model capabilities evolve, future systems with stronger reasoning or calibration may exhibit different debate dynamics. We also test a narrow range of team compositions and debate protocols; alternative designs, such as asynchronous deliberation, dynamic roles, or adversarial setups, remain open for exploration.

Third, while we propose three desiderata for effective debate (inclusive deliberation, rationale over assertion, and advancement of understanding), these criteria are tailored to logical reasoning tasks with verifiable ground truth. In domains where evidence is ambiguous or subjective, different effectiveness measures may be needed.

Finally, our experiments primarily measure debate outcomes through accuracy and state-transition analysis, but do not deeply investigate the linguistic or rhetorical mechanisms of persuasion. Understanding \emph{how} agents construct arguments, respond to counterarguments, or employ fallacies could further illuminate the conditions under which debates succeed or fail.


\balance
\bibliography{bibliography}

\appendix
\onecolumn
\section{Use of LLMs}
During the development of this paper, we use LLM Assistants in the following aspects: (i) Reference discovery: use the deep research tools from major providers to explore relevant work and literature. (ii) Code assistance: use coding agents to assist developing the code base of the current work. (iii) Grammar check: use LLMs to detect grammar errors in the drafty version of the paper, for better displaying our results.

Our experiments also involves testing the capacity of different models. This involves models from a few different major providers: OpenAi, Gemini, Qwen, and Bytedance. We didn't calculate the total token cost per provider, but an overall budget for all providers is around \$500.

\section{Outcome-level Quantitative Metrics}
\label{appendix:metrics}

\paragraph{Final Output Accuracy.}
By default we report \textbf{strict accuracy} over instances:
\[
\text{StrictAccuracy} = \frac{1}{N}\sum_{i=1}^{N} \mathbf{1}[\hat{y}_i = y_i],
\]
where $\hat{y}_i$ is the final predicted assignment for instance $i$ and $y_i$ is ground truth. 
We optionally report a \textbf{smooth} variant that averages per-player correctness within each instance:
\[
\text{SmoothAccuracy} = \frac{1}{N}\sum_{i=1}^{N}\frac{1}{P_i}\sum_{p=1}^{P_i}\mathbf{1}[\hat{y}_{i,p} = y_{i,p}].
\]

\paragraph{AUC-Strict-Accuracies.}
To summarize how strict correctness evolves during debate, we define the area-under-curve of strict accuracy across rounds:
\[
\text{AUC\_Strict} = \frac{1}{T}\sum_{t=1}^{T} \text{StrictAccuracy}(t),
\]
where $\text{StrictAccuracy}(t)$ equals $1$ if, at round $t$, the round-wise majority prediction matches ground truth for \emph{all} players in an instance, and $0$ otherwise. Values are in $[0,1]$; higher is earlier and more sustained correctness.

\paragraph{AUC-Smooth-Accuracies.}
Analogously, we aggregate the per-round smooth accuracy:
\[
\text{AUC\_Smooth} = \frac{1}{T}\sum_{t=1}^{T} \text{SmoothAccuracy}(t),
\]
where $\text{SmoothAccuracy}(t)$ is the proportion of players correctly predicted by the round-$t$ majority. Values are in $[0,1]$.

\paragraph{AUC-Agreement-All.}
We measure how often agents unanimously agree on player roles during the debate:
\[
\text{AUC\_Agree-All} = \frac{1}{T}\sum_{t=1}^{T}\text{AgreeAll}(t),
\]
with
\[
\text{AgreeAll}(t) = \frac{1}{N}\sum_{i=1}^{N}\frac{1}{P_i}\sum_{p=1}^{P_i}\mathbf{1}\big[\text{all agents agree player }p\text{ at round }t\big].
\]
Values are in $[0,1]$ and reflect unanimous player-level alignment over time.

\paragraph{AUC-Agreement-Major.}
We also measure majority agreement (at least half of agents agree) over time:
\[
\text{AUC\_Agree-Major} = \frac{1}{T}\sum_{t=1}^{T}\text{AgreeMajor}(t),
\]
with
\[
\text{AgreeMajor}(t) = \frac{1}{N}\sum_{i=1}^{N}\frac{1}{P_i}\sum_{p=1}^{P_i}\mathbf{1}\big[\max_{\ell}\#\{a:\hat{y}_{i,p,a,t}=\ell\}\ \ge \ \lceil A/2 \rceil\big],
\]
where $A$ is the number of agents. Values are in $[0,1]$; higher indicates earlier and more sustained majority alignment.

\section{Prompt Design of the Multi-Agent Debate System}
\label{app:prompt}
We now detail the prompts used for each phase described in Section~\ref{sec:method}.

\subsection{Prompt for Initial Proposal}
\begin{lstlisting}
    You are participating in a multi-agent debate about a Knight-Knave-Spy game. There are {num_player} other players in the game, each assigned a role of knight, knave, or spy. Each player will make a statement about themselves and other players. Besides those players, there is also a game manager who will provide you some hints. 

Game Rules:
- Knights always tell the truth.
- Knaves always lie.
- Spies can either tell the truth or lie.
- The hints from the game manager are always true.

Your task is to deduce the role of each player based on the statements and hints. You will be participating in a debate with other agents, and you can see the conversation history including your own previous responses and those of other agents.

The game info will be given in the following format:

---
Player name: string
Player statement: string
---
...
---
Message from the game manager: string (note this message is always true)

Your task is to deduce the role of each player, based on the statements and the hints. The result should be given in the following json format:
{
    "players": [
        {"name": "player_name", "role": "role"},
        ...
    ],
    "explanation": "string"
}

The players array contains objects with name and role strings. Include one entry for each player. Explanation is a string that contains the argument to derive the result.

Sample game info:
---
Player name: Violet
Player statement: Among all players, the number of spies is odd.
---
Player name: Uma
Player statement: Violet is lying.
---
Player name: Xavier
Player statement: Among Violet and I, there is exactly one knave.
---
Message from the game manager: I am the game manager and here is a hint for you: Among all players, there is exactly one spy.

Sample return:
{
    "players": [
        {"name": "Violet", "role": "knight"},
        {"name": "Uma", "role": "knave"},
        {"name": "Xavier", "role": "spy"}
    ],
    "explanation": "the argument to derive the result."
}

To fulfill the task, you should use the hints to deduce the truthfulness of the statements. You need to exhaust the possibility of the truthfulness of the statements and the role assignments, and either run into a contradiction which implies that the assumption is false, or reach a conclusion that is consistent with the hints. Rules for reasoning:
- Do not make extra assumptions beyond the game rules and hints. For example, if not mentioned explicitly, do not assume that there must be any particular roles among players.
- Do not conclude that the roles are not conclusive before you explicitly find two possibilities that are consistent with all the rules and hints, in which case you should mention both or all possibilities in the explanation.

Sample reasoning for the sample game info:
- The hint tells us that there is exactly one spy, so Violet is telling the truth. By the rule for the roles, he cannot be a knave.
- Since Violet is telling the truth, Uma must be lying. By the rule for the roles, she cannot be a knight.
- Assume Xavier is telling the truth, since we have deduced that Violet is not a knave, Xavier must be the knave himself, but the rule imposed that knaves always lie, so this is a contradiction. Therefore, Xavier must be lying.
- Since Uma and Xavier are both lying, and the hint says there is exactly one spy, one of them must be a knave.
- Assume Xavier is the knave, then since Violet cannot be a knave, Xavier is actually telling the truth, which contradicts with the rule that knaves always lie. Therefore, Xavier must be the spy, and Uma must be the knave.
- Since there is only one spy, and Xavier has taken the slot, Violet must be the knight. Therefore, the output is (remember to follow the format strictly):
{
    "players": [
        {"name": "Violet", "role": "knight"},
        {"name": "Uma", "role": "knave"},
        {"name": "Xavier", "role": "spy"}
    ],
    "explanation": "copy the above arguments here."
}

Keep your explanation having details but less than 100 words.

CRITICAL REQUIREMENTS:
- You MUST assign each player one of the three roles: "knight", "knave", or "spy"
- Do NOT use "unknown" or any other value - you must make a definitive choice for each player
- Base your decision on the game logic and evidence from the statements and hints
- If you're uncertain, choose the most likely role based on available evidence

Please follow strictly the format of the return, or the response will be rejected.
\end{lstlisting}

\subsection{Prompt for Player-by-player Debate Loop}

\textbf{Prompt for the Debate Phase}

\textit{Main Debate Prompt Template}

\begin{lstlisting}
    You are {agent_name} participating in a debate about {player_name}'s role in this Knight-Knaves-Spy game.

GAME INFORMATION:
{game.text_game}

CURRENT FOCUS: We are debating the role of {player_name}.{agents_context}{previous_context}

You can see the conversation history above, which includes:
- Your own previous responses (marked as your messages)
- Other agents' positions and reasoning
- The debate context and instructions

OTHER AVAILABLE AGENTS IN THIS DEBATE: {other_agents_list}

Your task is to:
1. Analyze the other agents' positions on {player_name}'s role from the conversation history
2. Decide which OTHER agents you agree with and which you disagree with (do NOT include yourself)
3. Provide reasoning for your agreements and disagreements
4. Make your final decision on {player_name}'s role

CRITICAL REQUIREMENTS:
- You MUST assign {player_name} one of the three roles: "knight", "knave", or "spy"
- Do NOT use "unknown" or any other value - you must make a definitive choice
- Do NOT consider other players' roles when making your decision (only consider {player_name})

Note: You can see your own previous responses in the conversation history, so you have natural self-awareness of your own position.

Return your response in JSON format:
{
    "player_role": "{player_name}",
    "role": "knight/knave/spy",
    "agree_with": ["other_agent_name1", "other_agent_name2"],
    "disagree_with": ["other_agent_name3"],
    "agree_reasoning": "Brief reasoning for agreements",
    "disagree_reasoning": "Brief reasoning for disagreements"
}
\end{lstlisting}

\textit{Current Agent Positions (agents\_context)}

\begin{lstlisting}
    CURRENT AGENT POSITIONS:
- {agent_name1} (YOU): {role}
  Reasoning: {explanation}
  Confidence: {confidence} (if enabled)
- {agent_name2}: {role}
  Reasoning: {explanation}
  Confidence: {confidence} (if enabled)
... (for each agent)
\end{lstlisting}

\textit{Previous Debate Rounds (previous\_context)}
\begin{lstlisting}
    PREVIOUS DEBATE ROUNDS:
Round {round_number} ({previous_player_name}):
  - {agent_name1} (YOU): {role}
    Reasoning: {explanation}
  - {agent_name2}: {role}
    Reasoning: {explanation}
  - CONSENSUS: {majority\_role}
... (for each previous round)
\end{lstlisting}

\textbf{Prompt for the Self-adjustment Phase}

\textit{Main Self-Adjustment Prompt Template}

\begin{lstlisting}
    You are {agent_name}. Based on the debate about {player_name}'s role, please provide your complete solution for ALL players.

GAME INFORMATION:
{game.text_game}

CURRENT FOCUS: We just finished debating {player_name}'s role.{debate_analysis}{previous_context}{latest_solutions_context}

You can see the conversation history above, which includes:
- Your own previous responses and solutions
- Other agents' positions and reasoning
- The debate arguments and agreements/disagreements
- Each agent's latest complete solution for all players

Based on the debate, please provide your self-adjustment solution on all players. 

CRITICAL REQUIREMENTS:
 - In this self-adjustment phase, you must provide a complete assignment of roles for all players, not just {player_name}.
 - Do not use "unknown" or any placeholder values - you must make a definitive choice for each player from "knight", "knave", or "spy".

Return your complete solution in JSON format:
{
    "players": [
        {"name": "player_name", "role": "role"},
        {"name": "another_player", "role": "role"},
        ...
    ],
    "explanation": "Your reasoning after considering the debate"
}

REMINDER: Include ALL players in the "players" array, each with their assigned role.
\end{lstlisting}

\textit{Current Debate Analysis (debate\_analysis)}
\begin{lstlisting}
    CURRENT DEBATE ANALYSIS:
- {agent_name1} (YOU): {role}
  Confidence: {confidence} (if enabled)
  Agrees with: {agent_names}
  Agree reasoning: {reasoning}
  Disagrees with: {agent_names}
  Disagree reasoning: {reasoning}
- {agent_name2}: {role}
  Confidence: {confidence} (if enabled)
  Agrees with: {agent_names}
  Agree reasoning: {reasoning}
  Disagrees with: {agent_names}
  Disagree reasoning: {reasoning}
... (for each agent in the debate)
\end{lstlisting}

\textit{Previous Debate Rounds (previous\_context)}
\begin{lstlisting}
    PREVIOUS DEBATE ROUNDS:
Round {round_number} ({previous_player_name}):
  Debate phase:
    - {agent_name1} (YOU): {role}
      Agrees with: {agent_names}
      Agree reasoning: {reasoning}
      Disagrees with: {agent_names}
      Disagree reasoning: {reasoning}
  Self-adjustment phase:
    - {agent_name1} (YOU): {role}
      Reasoning: {explanation}
  - CONSENSUS: {majority\_role}
... (for each previous round)
\end{lstlisting}

\textit{Latest Complete Solutions (latest\_solutions\_context)}
\begin{lstlisting}
    LATEST COMPLETE SOLUTIONS FROM EACH AGENT:
- {agent_name1} (YOU):
    {player1}: {role}
    {player2}: {role}
    {player3}: {role}
    Confidence: {confidence} (if enabled)
    Reasoning: {brief_explanation}
- {agent_name2}:
    {player1}: {role}
    {player2}: {role}
    {player3}: {role}
    Confidence: {confidence} (if enabled)
    Reasoning: {brief_explanation}
... (for each agent)
\end{lstlisting}

\subsection{Prompt for Final Decision}

\textbf{Final Discussion Prompt (for Agents)}
\begin{lstlisting}
    This is the FINAL DISCUSSION phase. You have access to the complete debate history.

GAME INFORMATION:
{game.text_game}{initial_summary}{debate_summary}

You can see the conversation history above, which includes:
- Your own responses throughout all phases (initial, debate, self-adjustment)
- Other agents' positions and reasoning from all phases
- The complete debate history and evolution of arguments

Now make your final decision for ALL players. You can reference the entire conversation history including your own responses and those of other agents.

CRITICAL REQUIREMENTS:
- You MUST assign each player one of the three roles: "knight", "knave", or "spy"
- Do NOT use "unknown" or any other value - you must make a definitive choice for each player

Return your final solution in JSON format:
{
    "players": [
        {"name": "player_name", "role": "role"},
        ...
    ],
    "explanation": "Your final comprehensive reasoning"
}
\end{lstlisting}

\textit{Initial Proposals Summary (initial\_summary)}
\begin{lstlisting}
    INITIAL PROPOSALS:
- {agent_name1}: {player_role_assignments}
  Reasoning: {explanation}
  Confidence: {confidence} (if enabled)

- {agent_name2}: {player_role_assignments}
  Reasoning: {explanation}
  Confidence: {confidence} (if enabled)

... (for each agent)
\end{lstlisting}

\textit{Debate Rounds Summary (debate\_summary)}
\begin{lstlisting}
    DEBATE ROUNDS AND SELF-ADJUSTMENT SUMMARY:
Round {round_number} ({player_name}):
  - {agent_name1}: {role}
    Confidence: {confidence} (if enabled)
    Agrees with: {agent_names}
    Agree reasoning: {reasoning}
    Disagrees with: {agent_names}
    Disagree reasoning: {reasoning}
  - {agent_name1} (self-adjustment): {player_role_assignments}
    Reasoning: {explanation}
    Confidence: {confidence} (if enabled)
  - CONSENSUS: {majority\_role}

... (for each round)
\end{lstlisting}

\textbf{Supervisor Prompt (for Final Decision if Majority Vote not Reached)}
\begin{lstlisting}
    You are a supervisor AI tasked with making the final decision in a multi-agent debate about a Knight-Knaves-Spy game.

GAME INFORMATION:
{game.text_game}

COMPLETE DEBATE HISTORY:

INITIAL PROPOSALS:
{proposal.agent_name} initially proposed: {proposal.player_role_assignments}
Their reasoning: {proposal.explanation}

DEBATE ROUNDS:
--- Round {round_number}: {player_name} ---
{response.agent_name} thought {player_name} is a {role}
  Agreed with: {agent_names} - Reasoning: {reasoning}
  Disagreed with: {agent_names} - Reasoning: {reasoning}
Reasoning: {explanation}

SUPERVISOR INSTRUCTIONS:
As the supervisor, you have access to the complete debate history. The agents have been unable to reach consensus, so you must make the final decision. Consider:

1. All initial proposals and their reasoning
2. The evolution of arguments through the debate rounds
3. The consistency and logic of each agent's reasoning
4. The overall coherence of the solution
5. Any patterns or insights that emerged during the debate

Make your decision based on the most logical and well-reasoned arguments you observed.

Return your response in the same JSON format:
{
    "players": [
        {"name": "player_name", "role": "role"},
        ...
    ],
    "explanation": "Your final decision with comprehensive reasoning based on the complete debate history"
}

IMPORTANT: Keep your explanation having details but less than 100 words.
\end{lstlisting}

\section{Prompt Used for DeepSeek Assessment in Figure~\ref{fig:mad-reasoning}}
\label{app:deepseek_prompt}
\begin{lstlisting}
    You are evaluating the logic soundness for an agent debate. There are {num_agents} agents in the debate. The debate is about a Knight-Knave-Spy game instance. The game setup is the following:
{{game_text}}

The debate focused on the role of the player {{player_name}}. There are three stages:

Stage 1: Initial proposal. Each agent propose a role deduction of this player, together with his reasoning.
Stage 2: Each agent decide to agree or disagree with other agents, and provide reasoning why he make such decisions.

Your task: to evaluate whether the agree and disagree reasons of the agent {{agent_name}} are sound or not. This helps us to evaluate how well the debate goes. To assist you, the correct solution of the game instance is:
{{ground_truth}}

In the user's input, the initial proposals of all agents and the agree and disagree info of the {{agent_name}} agent you are currently targeting will be provided. Your task is to evaluate whether these reasonings are solid enough to support the decision. Please follow the syllabus:
Rate 0: no applicable (i.e., no agree or disgree agents)
**For disagree reason**
- Rate 1: chaos. Show chaos in reasoning and behavior. For example, disagree with other agent having the same role deduction with self, but do not provide any logically sound argument.
- Rate 2: dangerous. Simply state the mismatch in role deduction as reason, or arguing that a correct reasoning from other agents is wrong.
- Rate 3: promising. Correctly point out the error from other agents as the disagree reason, but provide a wrong or gap argument why the other agents went wrong.
- Rate 4: solid. Correctly point out the error from other agents, and the argument why they are wrong is also solid.
**For agree reason**
- Rate 1: chaos. Show chaos in reasoning and behavior. For example, agree with other agent having the different role deduction with self, but do not provide any logically sound argument.
- Rate 2: dangerous. Simply state the match in role deduction as reason, without any further explanations. Suggesting the agent is simple compare the result rather than check the logic.
- Rate 3: promising. Correctly realize that the errors or limitations of self or other agents reasoning and behave accordingly (For example, realize self argument is wrong and hence agree with others), but either do not provide enough reasoning supports for this observation or still making wrong decisions (for example, agree with another wrong answer)
- Rate 4: solid. Correctly realize the errors or limitations, provide solid arguments, and behave correctedly (for example, realize error in reasoning but determine the final result is still correct, so decide to agree).

Please return your evaluation purely in the following json format:
{
    "agree_reasoning_soundness": int,
    "disagree_reasoning_soundness": int
}

Please note: 
- Focus on the logic soundness for your evaluation.
- Please strictly follow the response format.
\end{lstlisting}

\end{document}